# How to optimize neuroscience data utilization and experiment design for advancing brain models of visual and linguistic cognition?

**Greta Tuckute[1]** \* | **Dawn Finzi[2]** | **Eshed Margalit[2]** | **Jacob Yates[3]** | **Joel Zylberberg[4]** | **Alona Fyshe[5]** | **SueYeon Chung[6]** | **Evelina Fedorenko[1]** | **Nikolaus Kriegeskorte[7]** | **Kalanit Grill-Spector[2]** | **Kohitij Kar[4]** \*

[1]Massachusetts Institute of Technology
[2]Stanford University

[3]University of California - Berkeley

[4]York University
[5]University of Alberta
[6]New York University
[7]Columbia University

**Correspondence**
Kohitij Kar, Department of Biology, York University, Toronto, Ontario, M3J1P3, Canada
Email: k0h1t1j@yorku.ca

In recent years, neuroscience has made significant progress in building large-scale artificial neural network (ANN) models of brain activity and behavior. However, there is no consensus on the most efficient ways to collect data and design experiments to develop the next generation of models. This article explores the controversial opinions that have emerged on this topic in the domain of vision and language. Specifically, we address two critical points. First, we weigh the pros and cons of using qualitative insights from empirical results versus raw experimental data to train models. Second, we consider model-free (intuition-based) versus model-based approaches for data collection, specifically experimental design and stimulus selection, for optimal model development. Finally, we consider the challenges of developing a synergistic approach to experimental design and model building, including encouraging data and model sharing and the implications of iterative additions to existing models. The goal of the paper is to discuss decision points

**Abbreviations:** ANN: artificial neural network, AI: artificial intelligence, EEG: electroencephalogram, fMRI: functional magnetic resonance imaging

\*Equally contributing authors.

## 1 | INTRODUCTION

A core objective of neuroscience is to deepen our mechanistic understanding of the brain, which can be accomplished by developing computational models as explicit hypotheses. In recent years, many studies have focused on testing the accuracy of artificial neural network (ANN) models in matching primate behavioral and neural data, across the domains of vision (e.g., [1, 2, 3, 4, 5, 6]), audition (e.g., [7, 8, 9, 10]), and language processing (e.g., [11, 12, 13, 14, 15, 16, 17, 18], reviewed in [19]. In most of these cases, neural data have primarily been used to test whether ANN model responses match the brain responses using linear regression-based approaches (e.g., [2, 20]) or representational similarity analyses (e.g., [21]). However, recent studies have also demonstrated that primate neural data can be directly used (at various levels of abstraction) to finetune and optimize ANNs both to improve their match to brain responses as well as downstream task performance. For instance, simulating a primary visual cortex by matching the ANN front-end to empirically observed V1 properties improves robustness to image perturbations [22] while also better explaining higher-level visual cortical responses [23]. Similarly, [24] statistical properties of V1 activation patterns can serve as a teacher signal for training ANNs with resulting improvement in object recognition. More direct approaches include optimizing ANNs to align their representations to empirically observed neural responses in the macaque inferior temporal (IT) cortex, which has been shown to produce more primate-aligned, adversarially robust models of object recognition [25]. Related approaches have been proposed for language where language models can be modified to match brain activity better with resulting improvements in natural language processing tasks [26]. Another approach involves direct fine-tuning of language models on brain data, leading to improved brain predictivity performance, assessed through a different recording modality than the one used for fine-tuning [27]. These novel uses of neural data carry substantial promise for improving neural network performance and model alignment with primate neural and behavioral responses.

This article addresses two main challenges in the field of "NeuroAI" (intersection of neuroscience and artificial intelligence, AI; [28]). These two challenges focus on optimal model *development*, in contrast to model *evaluation* (e.g., [29]). **The first challenge concerns data usage in model development**. Fundamentally, do we believe it is the right time in neuroscience to use neural data to optimize models directly? Some neuroscientists [30, 31] (also see **Figure 1**) believe that we are still in the dark ages of neuroscience, which would suggest that more lay-of-the-land (exploratory) work needs to be conducted before we can begin collecting data at the granularity required to optimize model building directly. **The second challenge concerns future data collection efforts**. Should we rely on experimenter intuition derived from qualitative inferences generated by previous studies in the field, or should we use increasingly popular ANN-driven techniques such as the generation of "controversial" or "optimal stimuli" [32, 33, 34, 35, 36, 37] to guide experimental design? By addressing these challenges, we aim to develop more efficient ways of using neural data to build ANN models of the primate brain.

To set the stage for the two challenges, we first provide a brief review of the standard practices and recent advances in neuroscientific data collection, model development, and evaluation in visual and language neuroscience. We then elaborate on the challenges ("controversial axes") and evaluate the pros and cons associated with each axis. Furthermore, a key aim of the article is also to address how experimentalists and modelers could interact most productively going forward: For researchers who are in a position to collect data, how should they be using models to inform their experiments, and similarly for researchers who build models, how should they engage with experimentalists?

## 1.1 | Data and model interaction in neuroscience

The field of neuroscience has undergone a foundational shift with the emergence of task-optimized ANN models, offering computational accounts of neural processes. Although models have been central to scientific cycles of data collection and knowledge refinement for centuries, the capabilities afforded by the new class of ANN models differ from traditional approaches (e.g., ANN models are "stimulus computable" [38] and can provide quantitative predictions for arbitrary stimuli). In this section, we provide a brief review of interactions between data and models in neuroscience.

In vision research, early experiments by Hubel and Wiesel [39, 40] used microelectrodes to record the activity of individual neurons in the visual cortex of cats and monkeys. They found that these neurons responded selectively to specific visual features, such as particular orientations of light. Based on a combination of anatomical and physiological findings, Felleman and Van Essen [41] proposed that the visual cortex is a distributed hierarchical processing system. In addition, many studies have further discovered various aspects about these brain areas [42, 43, 44, 45, 46], generating inductive biases that form the basis of the development of the hierarchical model of visual processing, which proposes that visual information is processed in a series of stages, with simple features detected in early stages and more complex features detected in later stages. With advances in computational techniques and refinement of questions, specific information about functional topographies [47, 48, 46], recurrent processing [4, 49, 50], and other aspects of visual processing [51] has been incorporated in the architecture and computational motifs of the latest models of primate vision [52, 53, 54, 55, 56, 57, 58]. In addition, recent techniques have demonstrated using neural data to directly fit or regularize computational models [59, 25, 60].

In language research, studies of patients with brain damage from the late 19th century led to the identification of a set of frontal and temporal brain regions as being critical for language production and perception [61, 62]. These studies led to a family of descriptive models known as the "Classical model" or the "Wernicke-Lichtheim-Geschwind model" [63, 64, 65]. This family of models was later claimed obsolete due to, e.g., lack of definition of the brain regions described by these models (Broca's and Wernicke's areas) as well as the conflation of speech and linguistic processing [66, 67]. Later evidence from linguistics and neuroimaging led to the proposal of several descriptive models of language processing (e.g., [68, 69, 70, 71]) which coarsely tie brain regions to particular cognitive process ("syntactic structure building" or "lexical access"). These models offered intuitive explanations of language processing but were based on words rather than mathematical or computational terms and did not provide quantitative predictions of neural responses to language. Behavioral research in psycholinguistics led to computational models of some aspects of linguistic processing. Examples of such models are surprisal-based models that posit that the degree of contextual predictability modulates language processing difficulty (e.g., [72, 73]). However, most of these models were limited by fixed vocabularies, methodological challenges in representing features associated with model internals, and only instantiated one specific hypothesis (e.g., about surprisal modulating behavioral or neural responses). Modern language models [74, 75, 76], in contrast, can provide representations for any arbitrary linguistic input with its preceding context, capturing various linguistic regularities through high-dimensional vectors (e.g., [77, 78, 79]). These language models have provided the field with a chance to build quantitative models of neural processes for language. In addition, empirical evidence from neuroimaging further elucidated the brain regions implicated in language processing (the fronto-temporal language network; [80, 81, 82], reviewed in [83]), providing a clear target for modeling efforts [19]. Multiple studies have now demonstrated that these modern language models can explain neural responses in the language network with relatively high accuracy (e.g., [11, 12, 13, 14, 15, 16, 17, 84]).

## 1.2 | Advances in neuroscience data collection

Recent years have seen significant advances in neuroscience data collection, opening up new opportunities for model development using types and amounts of data that were previously unattainable. For behavioral data, online crowd-sourcing platforms have enabled the collection of large labeled behavioral datasets (e.g., [85, 86] at a scale that has previously been unattainable. Many labs have developed similar home-cage kiosks that facilitate high-throughput behavioral data collection in non-human primates [87, 88, 89]. Improved neural recording techniques now enable simultaneous measurements of hundreds or even thousands of neurons in awake, behaving animals across several levels of spatial and temporal specificity, such as multi-unit spiking activity [90, 91] or in-vivo calcium imaging [92]. Human neuroscience has also witnessed notable improvements in non-invasive fMRI data acquisition and analysis, including the use of high field strengths for acquisition (e.g., 7T), and analysis pipelines targeted towards modeling of event-related designs which provide fine-grained neural responses to each experimental trial [93]. Additionally, there has been a growing trend towards collecting massive amounts of data from a small number of individuals ("deep data" approach, e.g., [94, 95, 96, 97]. An example of such a dataset within the vision sciences is the Natural Scenes Dataset (NSD; [97]), which provides high-quality brain responses from a relatively low number of individuals to tens of thousands of images across diverse categories. The popularity of NSD demonstrates the hunger for reliable neural targets for model benchmarking at a fine grain across diverse stimuli, as opposed to data collected for specific hypotheses to broad stimulus categories (e.g., high vs. low contrast images or sentences vs. lists of words).

Further, direct brain perturbation strategies can be utilized to guide hypothesis testing (and related data collection) in both animals [49, 98, 99, 100] and humans [101, 102, 103]. Brain perturbations can leverage predictions from quantitative models to modulate brain activity according to a desired target [32, 33, 34, 37, 36]. Apart from the acquisition of neural responses to these coarse-grain brain perturbations [100, 104], the past decades have seen significant advances in more targeted genetic methods for collecting neural data in an even more targeted manner. Two key techniques are optogenetics [105] and chemogenetics [106], which allow researchers to selectively activate or inhibit specific neurons or circuits in the non-human primate brain, establishing causal relationships between neural activity and behavior. In summary, recent advances in data collection, from large-scale behavioral datasets via online platforms to sophisticated neural recording techniques and targeted neural perturbation techniques, hold immense potential for developing novel types of models and refining existing ones.

## 1.3 | Advances in computational brain model development

The last decades have seen significant advances in the development of computational models for neuroscience, particularly in the fields of vision and language. In the field of vision, the development of computational models has been motivated by empirical findings on receptive fields [39, 40] and hierarchical cortical processing [41], computationally formalized by Fukushima as the Neocognitron [107]. Improvements to this basic structure led to the development of the HMAX model [108]. A significant breakthrough was attained about a decade ago when it was discovered that convolutional neural networks that achieve human-level visual categorization performance quantitatively predict neural responses [2, 109] and representational structure[1] in high-level visual cortex in non-human primates and human primates, respectively. In addition, they also allow precise control of neural activity [32, 33, 34], making them promising tools for clinical translation. Similarly, in the field of language, models have evolved from n-gram [110, 111] and distributional semantics models (e.g., [112, 113, 114]) to Transformers (e.g., [115, 74, 75, 116]), which have shown great success in natural language processing tasks such as text generation, question-answering, translation, summarization, and language-based reasoning (e.g., [117, 118, 119, 120]). These model advancements provide a new set of explicit hypotheses about how the brain processes visual and linguistic information and constrain models of the brain by quantitatively predicting from the raw sensory inputs the resultant brain activations.

In this article, the primary class of models that we engage with (with respect to the data from Section 1.2) are those that are stimulus-computable, highly performant in the task/behavior of choice, and falsifiable (i.e., make explicit, testable predictions). In addition, we aim to have interpretable model components besides the input and output [121], which can provide insights into the neural mechanisms implicated in the behavior. By engaging with these types of models, we can better understand the relationship between neural activity and behavior and make progress in developing more accurate and interpretable models of brain function. In addition, we consider model development within two different neuroscientific domains: vision and language. Until now, most modeling efforts have targeted each domain separately, but we anticipate increasing focus on integrating these modeling efforts (see Section 1.4).

## 1.4 | Considering an integrative approach to developing computational models of visual and linguistic cognition

At first glance, computational models of visual and linguistic cognition might appear very different from one another. Why, then, might we want to consider these two seemingly different domains in discussions related to model development? We believe that shared insights between visual and linguistic models should represent a productive path for modeling efforts in less studied domains and for developing multi-componential models encompassing several perceptual/cognitive systems in a single, integrated model that links from one perceptual/cognitive process to another. For instance, one obvious point of link between vision and language systems is reading, which serves as a promising testing framework for models that integrate processing in the visual ventral stream with language processing in the fronto-temporal language areas. Another direction involves leveraging visual and linguistic modalities simultaneously to better explain responses in brain areas implicated in "higher-level" visual function or semantic processing through multi-modal models (e.g., [122, 123]). Finally, insights from one domain might help the other domain. For instance, linguistic models can help characterize and interpret visual models [124, 125]. Below, we briefly summarize the current commonalities and differences between computational modeling of visual and linguistic cognition.

## 1.5 | Commonalities and distinctions in computational modeling of visual and linguistic cognition

As mentioned above, several similarities and differences exist in approaches to modeling visual and linguistic processing. One major similarity is that task-optimized models tend to predict neuroscience data well in both visual and linguistic domains. For example, in the visual domain, models that perform the task of object recognition better (typically using ImageNet [85]) also better predict neural activity in areas of visual cortex involved in object perception [2, 1, 126] (but see [127]). In the linguistic domain, models that are better at predicting the upcoming word also better predict brain activity during language processing [13, 14, 128] (but see [129]). Another commonality is the presence of geometric phenomena in both visual and linguistic models. Specifically, both domains have shown evidence of "untangling" properties where semantically related information [130, 131] become more efficiently organized in high-dimensional space.

On the other hand, one difference in the research approaches when modeling vision and language is the availability and use of human versus animal models. Visual studies have often leveraged the high temporal and spatial specificity available from invasive studies done in non-human primates, while linguistic studies rely primarily on human participants using non-invasive recording modalities (and coarser grain of brain measurements). Another difference is the level of domain knowledge required for modeling. In the visual domain, there is a wealth of knowledge about the structure and function of the visual system [132] that can be leveraged to inform models [127]. In contrast, there is less well-established domain knowledge in language processing, making the development of models with qualitative priors more challenging.

Below, we attempt to tailor our arguments to vision and language processing studies independently, wherever possible.

## 2 | THE CONTROVERSIAL AXES

As neuroscience continues to evolve rapidly (under limited resources [133]), researchers need to adapt their strategies to make the most of existing data, design effective experiments, and develop models that accurately capture the complexity of the brain. Addressing these high-level challenges produces certain issues (denoted as "controversial axes" in this article) that warrant careful consideration.

Below, we elaborate on the specific challenges and controversial aspects that formed the basis of our discussions during the Generative Adversarial Collaborations (GAC) session of the Conference on Cognitive Computational Neuro-science [134], 2022, structured as follows: Each sub-section opens with a "controversial" claim, then proceeds to an explanation of the approach proposed by the claim. Each sub-section provides supporting arguments for the proposed approach and concludes with a box that outlines counterarguments or key considerations to keep in mind.

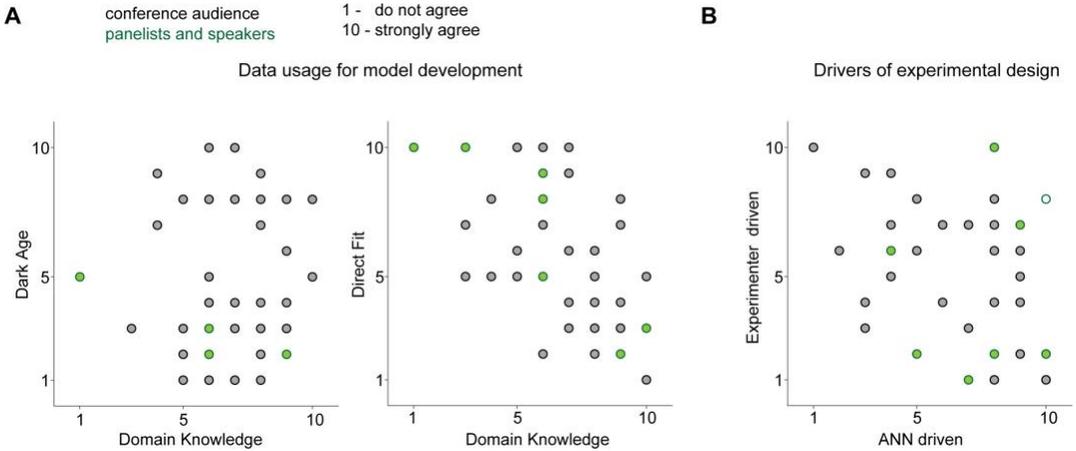

**FIGURE 1** Audience (n=35) and GAC participant (n=10) poll results. Each dot shows the rating provided by the conference audience (in black) and the GAC panelists and speakers (in green) for two of the three positions. 1 - Do not agree, 10 - strongly agree. **A**. Answers to the following three opinions. **Direct Fit:** "Experimental data (not insights from "textbooks") should be used to directly train artificial neural network models of brain activity and behavior." **Domain Knowledge:** "Qualitative insights (and not experimental data) from previous and existing experimental results should be used (e.g., as inductive biases) to design artificial neural network models of brain activity and behavior." **Dark Age:** "We are still in the dark ages of neuroscience, and more lay-of-the-land type ("fundamental") neuroscience work needs to be conducted before we start collecting data at a grain leveraged for building artificial neural networks of brain activity and behavior." The green-shaded region highlights a mismatch between the ratings from the audience and the GAC speakers. Unlike the speakers, the audience felt that neuroscience was still very much in its early days, and data collection optimized for model development might be a premature idea. **B.** Answers to the following two opinions. **ANN-driven:** "Experimental design should be based on artificial neural network models that are predictive of brain activity and/or behavior (for designing and optimizing experimental paradigms using, e.g., 'controversial'/'optimal' stimuli)." **Experimenter-intuition driven:** "Experimental design should be based on neuroscientists' intuition derived from qualitative inferences generated by previous studies in the field (e.g., V1 cells like oriented bars, let's find out what V2 cells like)."

## 2.1 | How do we optimally utilize neuroscience data for model development?

### 2.1.1 | Qualitative insights (not raw experimental data) from experiments should be utilized to design ANN models of brain activity and behavior.

**Qualitative insights**

In neuroscience, existing assumptions and knowledge about a brain area or specific mechanisms can be used to build ANN models without necessarily engaging with all the experimental data collected to develop this knowledge base. This approach incorporates inductive biases based on qualitative insights from existing experimental results. Examples of these insights include the hierarchical processing of information in the visual cortex [135, 46, 42, 47, 43], recurrent processing [4, 49, 136], nonlinear properties of neural activity, spatial (e.g., retinotopy, [137, 138, 139, 140]) and functional specialization (e.g., face-selective neurons) [141, 47], short and long time-scale adaptation [142, 143] among others. Arguably, the success of the current CNNs as models of primate vision [38, 144] can be attributed to such model architectures being influenced by decades of inductive biases from the primate visual neuroscience and psychology literature (see discussions in [127]). However, a recent study [6] has demonstrated that model architecture does not significantly alter its neural predictivity, emphasizing for the need of stronger constraints to leverage architectural guidance in model development. Consistent with this observation, recent progress in the development of large-scale language models [75, 74, 76, 116], which also serve as state-of-the-art models of language processing in humans (e.g., [13]) also arguably demonstrates that successful models do not always require insights from neuroscience.

**Arguments supporting this approach**

Employing a scientific approach based on qualitative insights offers several advantages. Qualitative insights allow for the abstraction of general concepts into the form of inductive biases within ANN models. Importantly, such concepts lack idiosyncrasies of any given neuroscience dataset and/or recording modality. Incorporating general concepts of perceptual/cognitive functions explicitly in models in the form of inductive biases is a direct way of testing if that factor is important to account for the relevant neuroscientific data, improving our "understanding" of the brain. In a regime where data are limited, qualitative insights are the only resource to develop and advance models. Qualitative insights, e.g., recurrent processing, can be instantiated in a series of (very different) models [52, 53] and can subsequently be benchmarked against empirical data and compared to models developed using other insights (e.g., feedforward processing with constrained topographic organization, [56, 55, 145]). The summary of these outcomes can drive broad trajectories of future models.

> **Things to keep in mind (for the approach in Section 2.1.1)**
>
> The decision process of which insights to incorporate into a model is not trivial, and it is important to ensure that the chosen inductive biases are causally important and not merely epiphenomenal. There is an ongoing debate about the relative merits of direct fit versus inductive biases in model development (e.g., [146, 147]).

> **Definition of terms.**
>
> **Making progress in neuroscience:** Building falsifiable, quantitatively accurate brain models leads to progress in visual and linguistic cognition.
>
> **Neuroscience data:** Neuroscience data includes both neural data, such as data obtained from electrophysiology or brain imaging techniques, and behavioral data, such as data obtained from perceptual or psychological experiments.
>
> **Benchmarks:** A quantitative measure of how well a brain model can predict empirical neuroscience data.
>
> **Inductive bias:** Inductive bias refers to the set of assumptions or prior knowledge built into the design of a machine learning algorithm or artificial neural network (ANN). In neuroscience, inductive bias can be considered the set of constraints or assumptions based on our understanding of the brain and how it processes information.
>
> **Direct fit:** Direct fit refers to fitting a model to empirically measured neuroscience data by directly optimizing the model's parameters to minimize the difference between the model's predictions and the empirical neuroscience data.
>
> **Animal (in-vivo) model:** Animal (in-vivo) model refers to using living animals (e.g., rhesus macaques) as a model system to study neural function and behavior to make inferences about neurobehavioral aspects in another species (e.g., humans).
>
> **Computational (in-silico) model:** Computational (in-silico) model refers to using computer simulations (e.g., deep convolutional neural networks, large language Transformer models) of behavior and neural function to make inferences about neurobehavioral aspects in another species (e.g., humans).
>
> **Prediction:** Prediction refers to the ability to quantitatively forecast or anticipate neural activity in a brain region or behavior in a specific task.
>
> **Control:** Control refers to the ability to perform closed-loop manipulation of neural activity and behavior through optogenetic, electromagnetic, chemogenetic, or pharmacological interventions.
>
> **Understanding:** Understanding in this context refers to the ability to create falsifiable computational models to make accurate predictions, provide a basis for control, and explain underlying neural mechanisms via model components that are interpretable to experimenters.

## 2.1.2 | Experimental data (not qualitative insights from "textbooks") should be used to directly train ANN models of brain activity and behavior.

**Quantitative insights**

Using large-scale behavioral and neural experimental data to train ANN models directly represents an emerging approach in neuroscience research. In this approach, behavioral or neural signals are a rich source of information to learn better algorithms from, either directly or indirectly. The most straightforward approach is to train models to directly predict behavioral or neural responses. Another approach is to leverage neuroscientific data indirectly, e.g., as part of the loss function for models trained to perform a normative task (e.g., [148, 149]). In neuroscience, there has been a long tradition of using statistical methods, such as Poisson likelihood, to fit likelihood-based models directly to neural spike trains, and modern ANNs fit naturally in this framework. The current best models of early visual areas, such as the retina and primary visual cortex in mice and monkeys, have been developed using this approach [150]. Various other recent studies [25, 151, 60] have demonstrated the effectiveness of directly optimizing models on experimental data for higher visual cortices. Consideration of the various forms in which neuroscientific data can be directly integrated with a model's training regime, e.g., in the form of abstractions of data like representational dissimilarity matrices [152], can lead to better models of behavior and brain activity.

The tradition of incorporating quantitative insights is less prevalent in language compared to vision, but studies

[153, 154, 27, 155] have made efforts to leverage neuroscientific data for language-based tasks such as sentiment analysis or relation detection.

**Arguments supporting this approach**

Data-driven ANNs, trained using specific neural population activity or behavioral patterns, offer a direct way to develop models of targeted behaviors or brain areas. Direct fit to neuroscientific data can provide shortcuts to a more accurate representation of the mechanisms for the behavior of interest or response patterns in the targeted brain area (e.g., [25, 60, 156]). This approach can provide a complementary strategy of model development that does not rely on computational search for the optimal architecture, loss function, and task to develop a goal-directed ANN that, in turn, might have internals with brain-aligned representations. Directly training on experimental data may lead to the emergence of the exact implicit mechanisms one might impose as an inductive bias arising from qualitative insights. If such common mechanisms emerge naturally through direct fitting, it would provide strong support for using this strategy in the first place, bridging the gap between data-driven and theory-driven approaches. Although data limitations exist when fitting directly to neuroscientific data, combining data across multiple experimental sessions can help overcome this challenge [157]. Compared to the inferences abstracted (and therefore compressed) as qualitative insights, direct fitting, when performed in a data-rich regime, is likely to be free of the biases introduced by the cognitive capacity of a human experimenter. In other words, this approach helps to avoid experimenter bias in the interpretation of results, as feeding data directly into the model may expose it to factors that are not immediately interpretable to a human experimenter.

> **Things to keep in mind (for the approach in Section 2.1.2)**
>
> The success of direct fitting is strongly dependent on the scale and richness of the available neuroscientific data. The nature of the data used for training significantly influences a model's predictive capabilities. For example, training predominantly on static images like those from ImageNet limits the ability to predict dynamic stimuli, such as naturalistic motion. This highlights the need for diversifying training data to enhance the scope and accuracy of model predictions. Furthermore, even with ample data, there is inherent tension between fitting large expressive models and smaller, biologically constrained models that are more prone to local minima. Integrating various types of neuroscientific data, such as single units and voxels, across different species on the same stimulus set could provide invaluable insights. If a single model can explain data across these modalities and species, it might reveal a universal principle of brain organization (e.g., [158]). Conversely, if adjustments are needed to accommodate humans and non-human primates, it could offer insights into how these brain systems differ. Therefore, assessing the type, diversity, and sufficiency of data is crucial. Developing theoretical tools for this assessment will significantly determine the suitability of the approach for model development, ensuring that the models not only fit well but also are representative and capable of making accurate predictions across various stimuli and neuroscientific contexts.

## 2.2 | How do we optimally design future experiments?

### 2.2.1 | Experimental design should be based on neuroscientists' intuition derived from qualitative inferences generated by previous studies in the field

**Model-free experimental design**

Traditionally, the field of neuroscience has relied heavily on both theoretical and incremental experimental work to guide experimental design and the selection of appropriate experimental stimuli. In the current work, we define "model-free" as designing experiments by choosing the experimental stimuli based on the experimenters' intuitions about the outcome ("Brain region A responds more to images of curved surfaces" or "Brain region B responds more to ambiguous words than unambiguous words"), without the use of an explicit computational model. Thus, model-free experiments only generate qualitative, descriptive predictions. In principle, there is also an implicit model leveraged in this case. But the model essentially runs in the experimenters' heads instead of predictions derived in-silico ("model-based"), making quantification of progress intractable.

In vision neuroscience, model-free approaches broadly fall into three categories. First are hand-constructed stimuli based on prior experimental evidence and experimenter intuition. These stimuli are often simple geometric shapes or patterns designed to isolate specific visual features. For example, Hubel and Wiesel [39, 40] used oriented bars and edges to study receptive fields in primary visual cortex, while Pasupathy and Connor [159] employed systematically varied shape contours to investigate shape encoding in area V4. Second are stimuli that are amenable to linear systems analysis or another systems identification approach [160, 161, 162]. These typically involve using white noise (for reverse correlation techniques) or other statistically controlled stimuli (e.g., drifting sinusoidal gratings) to characterize neural responses. Third, experimenters often use natural stimuli (either fully naturalistic, e.g., [163, 164], or augmented to represent rare stimuli occurrences, e.g., [165] or constructed to contain the statistical structure of natural images e.g., Najemnik and Geisler (2005) [166]. In language neuroscience, "traditional" (i.e., model-free) approaches for deriving stimuli for experiments broadly fall into two categories: i) hand-constructed stimuli based on psycholinguistic evidence, theoretical constructs, and/or experimenter's intuition (e.g., [167, 168, 169, 170, 71]) often organized in minimal pairs where sets of sentences are designed to differ minimally from one another (such as a verb that is correctly/incorrectly conjugated), ii) stimuli sampled from naturalistic corpora (either fully naturalistic as, e.g., [122] or 'deceptively' naturalistic as for instance [171, 172] where stimuli are constructed to contain an over-representation of rare stimuli occurrences than would otherwise be embedded in fully naturalistic stimuli distributions).

**Arguments supporting this approach**

Model-free approaches for deriving experimental stimuli can, in some cases, provide several advantages, including interpretability, explicit control of confounding factors, and inclusion of infrequently occurring phenomena (e.g., [171, 172, 173, 174], as elaborated below.

One key advantage is interpretability: Experimenters can choose stimuli directly linked to a descriptive, model-free hypothesis being tested, making it tractable to interpret results and draw conclusions, albeit limited by the boundaries of what humans can explain and put into words. Hence, if the experimental outcome falls within the possible scenarios hypothesized by the experimenter, the experiment provides a degree of abstraction in understanding the link between stimuli and outcome in the brain region. Interpretability, in combination with the experimenter's ability to reason about potential confounds and stimulus properties that are uninteresting to the hypothesis being tested, allows for even more targeted investigations of the phenomenon of interest. Another key advantage of experimenter-derived stimuli is the ability to include rare phenomena. Within language, some linguistic phenomena are rare albeit useful for investigating mechanisms of language processing [175, 176]. Language models are trained to generate the most plausible next word given the preceding context [75]. If stimuli are derived directly from these models, they will likely miss out on these less common phenomena. In contrast to experimenter-designed stimuli, with natural stimuli, a strong argument can be made that they should be used because they contain the relevant input statistics for the system that evolved to process them [177].

> **Things to keep in mind (for the approach in Section 2.2.1)**
>
> While model-free stimulus selection is often based on the experimenter's targeted hypotheses, it also encompasses a wider range of choices that typically may not be considered as hypotheses. A prominent example is the use of stimuli with statistical properties of natural sensory input [178], which can also be considered an experimenter's intuition-driven set of stimuli. However, it is an intuition grounded in the assumption that an accurate brain model should similarly process naturalistic input as the primate brain (e.g., [179]). Stimuli derived for neuroscience experiments based on strong top-down assumptions of the experimenter often come with strong expectations of finding parallels between the hypothesized representations and organization of the brain. For example, within vision, it was hypothesized that motion detection in the primate brain involves temporal delays due to intrinsic differences between fast (magnocellular stream) and slow synaptic transmission (parvocellular stream). However, despite strong expectations of finding a relationship between the expected temporal delay circuits in the brain and motion perception, the empirical evidence has been questionable [180, 181]. Similarly, within language, a large body of theoretical work focused on the distinction between syntax (form) and semantics (meaning), and the expectation was a parallel organization in the brain. Later studies showed a lack of empirical support for this distinction [182, 183, 71, 184]. These two examples demonstrate that neural populations might be tuned to many (possibly interacting) dimensions related to the stimulus, which are not immediately intuitive or interpretable to the experimenter.

### 2.2.2 | Experimental design should be based on artificial neural network models that predict brain activity and/or behavior

ANN models can be effectively used in designing and optimizing experimental paradigms, for instance, through "controversial" or "optimal" experimental stimuli. We define a "model-based" approach as one where experiments (and stimuli) are designed using explicit computational models that offer hypotheses in the form of quantitative predictions about the experimental outcome. In its most conservative form, such a framework posits that one should not design an experiment without explicit, quantitative predictions from a computational model. It is important to note that by model-based stimulus selection, we do not imply using sophisticated machine learning-based tools (e.g., generative models) to derive stimuli for experiments that are otherwise based on an experimenter's intuitive hypotheses. In a model-based approach, the computational model itself serves as the hypothesis. These models/hypotheses are stimulus-computable and make precise predictions for each stimulus. This approach is similar in principle but distinct from stimuli generation strategies used in previous studies based simpler, non image-computable models [159, 185].

To make the process of falsifying or validating the models more efficient, the models themselves can be used to generate stimuli: for example, models can be used to generate "optimal" stimuli, stimuli that predict a certain activation level in a location, e.g., maximal activation, in a group of neurons or a brain region [32, 33, 34, 186, 37, 36]. Another example is to generate "controversial" stimuli, stimuli for which models produce maximally distinct predicted responses, to test discrepancies between models and humans [187, 35]. These methods for "optimal" or "controversial" stimuli can be leveraged in real-time, providing substantial potential for efficient model evaluation, albeit with technical complexity [188, 189, 190, 191, 192].

Looking ahead, these methods might be instantiated in large-scale data collection efforts (e.g., The Natural Scenes Dataset [97]). Instead of human experimenters selecting the stimuli (e.g., choosing all images in the ImageNet dataset), one could envision collecting fully optimized datasets based on methods such as controversial stimuli (or more advanced methods), for more efficient probing of competing models.

ANN models mimicking the brain at multiple levels, also now referred to as SMART models (Sensory computable Mechanistic Architecturally Referenced Testable models [127]), show promising applications in various medical fields. In brain-computer interfaces, these models could enable complex stimulation patterns for visual areas [98], potentially recreating visual experiences for the visually impaired. For mental health interventions, SMART models might guide precise brain activity modulation, offering new treatment avenues beyond traditional pharmaceuticals and invasive probes. For instance, they could inform the creation of specific light patterns delivered through the retina to influence deep brain neuronal populations, potentially alleviating conditions like anxiety, depression, or other sensory differences (see [193]). The development of personalized SMART models, tailored to individual neurological and behavioral profiles, could transform clinical interventions. As these models become more comprehensive, integrating cellular and molecular components, they could open the door to powerful interventional tools at the molecular and genetic levels. Hence, ANNs ability to predict neural and behavioral data could guide the development of medical treatments, interventions, and devices like visual or language prostheses while also informing the types of experimental data to collect (i.e., experimental design choices) for effective medical advancements.

**Arguments supporting this approach**

Model-based approaches for determining experimental design (including but not limited to stimuli) can provide several advantages, including efficiency and a computationally tractable way of expanding the hypothesis space. The space of all possible stimuli that one might use as experimental stimuli is infinite. Given that the models serve as the hypothesis being tested, one can leverage their predictions to efficiently select stimuli that will most efficiently discriminate among them. This is in sharp contrast to an approach where stimuli for model evaluation are either selected randomly or based on the experimenters' intuition. Experimenter-derived stimuli often come with strong assumptions about the experimental outcome, while relying on models that instantiate these expected outcomes allows us to generate stimuli and associated predictions that are critical in efficiently falsifying such models and testing alternative theories. A model-based approach can help design experiments that allow us to test an expanded space of hypotheses in a more tractable manner than a model-free approach. For instance, using a model-based approach, Kar et al. (2019) [4] discovered specific images where primates outperformed the feedforward convolutional neural networks during object discrimination tasks, thereby exposing a large explanatory gap. This motivated the design of recurrent convolutional neural networks [4, 52, 53] aimed at specifically bridging this explanatory gap between primates and models. In summary, the model-based experimental design provides a tractable method for directly falsifying the model via its predictions, leading to a virtuous cycle of model falsification and improvement.

---

**Things to keep in mind (for the approach in Section 2.2.2)**

Inferences based on model-based experimental approaches are limited by the existing biases, idiosyncrasies, and/or discrepancies of the current generation of models in the stimulus selection process. These biases might be challenging to identify due to the black box nature (lack of human interpretability; [121]) of most modern ANNs. Using existing classes of models to generate stimuli with subsequent model validation might increase our chances of validating the models instead of discovering novel, potentially more accurate models. Additionally, by using models to close the experimental loop, we may be overfitting conditions where artificial and biological networks are aligned (e.g., processing of images) at the cost of ignoring entire dimensions of relevance for vision (e.g., time). Finally, the utility of a model-based versus model-free experimental design depends on the "maturity" of the research domain: In early stages of exploring a novel brain area or process, model-free experiments can serve as a way to obtain initial insights, particularly when the domain is not sufficiently defined to be represented by specific design choices of a model. When the initial groundwork is done, these insights can be operationalized as computational models for further investigations and targeted experiments.

# 3 | GOING FORWARD: MAKING, MEASURING AND UNDERSTANDING PROGRESS

In this concluding section, we discuss three issues related to optimal model development going forward. First, we discuss how to foster progress through data and model sharing. Next, we discuss how to measure progress through model evaluation. Finally, we discuss how to understand progress through theory and how theory can aid in developing better models.

## 3.1 | Fostering progress: Data and models

In this article, we have highlighted and discussed the approaches for optimal model development encompassing "traditional" experimental methods and more recent model-based techniques. However, a key challenge in enhancing collaboration between experimentalists and model developers lies in the sharing of data and models. To foster an environment conducive to data and model sharing, it is necessary to address technical, ethical, and policy-related issues. Technically, we should develop robust, secure, and user-friendly data infrastructures that support seamless sharing and collaboration, possibly utilizing advanced cloud technologies and decentralized systems. Progress has been made with platforms such as Neurodata Without Borders[1], OpenNeuro[2], and Distributed Archives for Neurophysiology Data Integration[3]. Regarding policy, fostering an academic culture and funding ecosystem that values and rewards data sharing will be critical. This approach involves implementing strong incentives around data ownership, access, and use for individual researchers and institutions to share their work. In addition, one must also carefully consider how data sharing for model development interacts with model evaluation. Certain datasets need to be stored privately (and regularly updated) by integrative benchmarking platforms (like Brain-Score, see Section 3.2) to ensure that the models are not directly fit on the test data. Finally, from an ethical perspective, it is critical to ensure privacy, confidentiality, and informed consent, especially when dealing with sensitive data, by establishing rigorous institutional protocols and legal agreements.

## 3.2 | Measuring progress: Model evaluation

This article has focused on model *development*, and not *evaluation*. However, development and evaluation are inherently connected, and a key question is whether model development will change as model evaluation changes. Current models are being developed and evaluated primarily using prediction-based measures such as regression [2]. These measures typically quantify the explained variance between the model's predictions and the measured responses (neural activity and behavior), corrected for noise in the data. Prediction-based metrics can also be leveraged for neural control where model-based predictions are used to identify stimuli aimed at eliciting a particular response in neurons or brain regions [33, 36], providing a more robust test of the model's predictions.

Another avenue of evaluation includes metrics that take into account the geometry of neural populations or ANN unit activations, such as representational similarity analysis (RSA) [194, 195], and centered kernel alignment (CKA) [196]. RSA involves constructing symmetric matrices that capture the pairwise distances between neural activity patterns for a set of test images, both for the measured brain responses and the model's predicted responses, and then comparing these matrices using a similarity metric such as Pearson correlation. CKA, on the other hand, is a similarity measure that is invariant to rotation and isotropic scaling but not to all linear transformations, providing an alternative way to assess the similarity between model and brain population representations. More recent approaches within geometric measures include manifold-based metrics, which allow for quantification of manifold dimensionality and separability of representations in high-dimensional neural and artificial populations [130]. Other recent geometry-based approaches include generalized shape metrics [197], pointwise representational similarity [198], or soft matching distance [199]. New evaluation measures are also being developed. For example, direct interfacing of brain responses into models [200] with subsequent evaluation of task performance represents an exciting direction within model evaluation. Another example is the direct comparison of responses to spoken language from the auditory brain stem recorded via electroencephalogram (EEG) and representations from generative audio networks [201], bypassing the transformations (e.g., via regression) that are typically used to evaluate the similarity between brains and models. Finally, several platforms now offer integrative benchmarking, which includes evaluation across a broad and diverse range of models, datasets, and metrics. Examples of such platforms are Brain-score[4] [109], Algonauts[5] [202, 203], and Sensorium[6] [204] which all aim to provide a more comprehensive picture of the computational models' alignment with the biological brain irrespective of the idiosyncrasies associated with smaller scale evaluations.

---

[1] https://www.nwb.org/
[2] https://openneuro.org/
[3] https://dandiarchive.org/

## 3.3 | Understanding progress: Linking models with theory

In discussions about model development, it is not uncommon to note diverse opinions about the notion that "we need more theory". In this section, we briefly discuss what we consider a theory and how linking model development with theory can lead to a fruitful cycle of model development.

A scientific theory (in neuroscience and general) synthesizes evidence, rationale, and ideas that help us explain empirical observations. Theories provide a high-level understanding of the phenomena under study. Theories are not directly testable but guide the development of computational models. They are tools by which scientists can:

1. Make sense of the observations, which typically involve mathematical formulations and abstractions of raw data using various statistical tools to understand the underlying patterns and structures (e.g., dimensionality estimates, manifolds, identifying specific spatiotemporal functional topographies).
2. Make predictions, which can be achieved by instantiating the theories as explicit computational models (like a trained neural network). These models allow for falsifiable predictions.

Given this broad definition of theory, we believe that theoretical work can be leveraged at various stages of model development, experiment design, and data collection. We emphasize that we view computational models as instantiations of theories, not theories per se: We view computational models as instantiations of theories. Some instances of the utility of theory are discussed below:

**When generating new hypotheses:** A robust theory can suggest novel and testable hypotheses. It can guide the instantiation of those hypotheses as explicit computational models. These models can then make predictions that explore new empirical data points that might have remained unexplored, ideally to falsify or discriminate amongst those same models.

**When integrating and interpreting data:** Theories can help in the mathematical formulation and abstraction of raw data, allowing scientists to discern patterns and structures that are not immediately obvious. The theoretical framework can inform which statistical tools to apply and how to interpret their outputs. For instance, it has been empirically observed that classification accuracy of object category improves along the ventral visual hierarchy [90] of the primate brain and the hierarchies in the deep convolutional neural networks. Recent theoretical work [130] has explained how the changes in the geometry of the neural-population responses to an object category ("object manifold") can be quantitatively summarized with "classification capacities" and how the classification capacity-based estimates of the manifolds' radius, dimensionality, and inter-manifold correlations, can further explain the functional roles of different parts of this hierarchy.

---

[4]https://www.brain-score.org/
[5]http://algonauts.csail.mit.edu/
[6]https://www.sensorium-competition.net/

**In refining experimental methods:** Theories can inform the design of experiments by predicting what conditions are most likely to yield informative results. This can optimize resource use and direct experimental efforts more efficiently.

**For iterative model-experiment cycles:** Theory-driven models can be used to generate predictions that are tested experimentally. The results of these experiments can, in turn, refine the theory, leading to a virtuous cycle of improvement in both theoretical understanding and model precision. Therefore, the mode of leveraging theory is likely cyclical.

## 4 | CONCLUDING REMARKS

In the rapidly evolving field of NeuroAI, the integration of neuroscience data with computational model development offers exciting and promising directions in the modeling of brain function, yet it also presents several challenges and decision points. In this paper, we focused on i) optimal usage of neuroscientific data in model development, contrasting raw experimental data with qualitative insights, and ii) optimal data collection efforts, contrasting model-based versus model-free approaches to experimental design. The main scope of the paper is to formulate and discuss the different opinions (at each end of the spectrum) on the role of data usage and experimental design in model development. In doing so, we hope that the readers will recognize the assumptions and considerations underlying these decisions.

### References

[1] Khaligh-Razavi SM, Kriegeskorte N. Deep supervised, but not unsupervised, models may explain it cortical representation. PLoS Computational Biology 2014 Nov;10(11):e1003915. https://dx.plos.org/10.1371/journal.pcbi.1003915.

[2] Yamins DLK, Hong H, Cadieu CF, Solomon EA, Seibert D, DiCarlo JJ. Performance-optimized hierarchical models predict neural responses in higher visual cortex. Proceedings of the National Academy of Sciences 2014 Jun;111(23):8619–8624. http://www.pnas.org/cgi/doi/10.1073/pnas.1403112111.

[3] Güçlü U, Thielen J, Hanke M, van Gerven MAJ. Brains on beats. In: Advances in Neural Information Processing Systems 29 (NIPS 2016); 2016. p. 2109–2117.

[4] Kar K, Kubilius J, Schmidt K, Issa EB, DiCarlo JJ. Evidence that recurrent circuits are critical to the ventral stream's execution of core object recognition behavior. Nature Neuroscience 2019 Jun;22(6):974–983.

[5] Mineault P, Bakhtiari S, Richards B, Pack C. Your head is there to move you around: Goal-driven models of the primate dorsal pathway. Advances in Neural Information Processing Systems 2021;34:28757–28771.

[6] Conwell C, Prince JS, Alvarez GA, Konkle T, Large-Scale Benchmarking of Diverse Artificial Vision Models in Prediction of 7T Human Neuroimaging Data. bioRxiv; 2022. https://www.biorxiv.org/content/10.1101/2022.03.28.485868v1, pages: 2022.03.28.485868 Section: New Results.

[7] Kell AJE, Yamins DLK, Shook EN, Norman-Haignere SV, McDermott JH. A task-optimized neural network replicates human auditory behavior, predicts brain responses, and reveals a cortical processing hierarchy. Neuron 2018 May;98(3):630–644.e16. https://linkinghub.elsevier.com/retrieve/pii/S0896627318302502.

[8] Millet J, Caucheteux C, Orhan P, Boubenec Y, Gramfort A, Dunbar E, et al. Toward a realistic model of speech processing in the brain with self-supervised learning. In: Advances in Neural Information Processing Systems 35 (NeurIPS 2022); 2022. .

[9] Vaidya AR, Jain S, Huth AG. Self-supervised models of audio effectively explain human cortical responses to speech. In: Proceedings of the International Conference on Machine Learning (ICML); 2022. https://doi.org/10.48550/arXiv.2205.14252, accepted on 27 May 2022.

[10] Tuckute G, Feather J, Boebinger D, McDermott JH. Many but not all deep neural network audio models capture brain responses and exhibit correspondence between model stages and brain regions. PLOS Biology 2023 12;21(12):1–70. https://doi.org/10.1371/journal.pbio.3002366.

[11] Jain S, Huth A. Incorporating context into language encoding models for fMRI. Advances in neural information processing systems 2018;31.

[12] Gauthier J, Levy R. Linking artificial and human neural representations of language. In: Proceedings of the 2019 Conference on Empirical Methods in Natural Language Processing and the 9th International Joint Conference on Natural Language Processing (EMNLP-IJCNLP) Hong Kong, China: Association for Computational Linguistics; 2019. p. 529–539. https://www.aclweb.org/anthology/D19-1050.

[13] Schrimpf M, Blank IA, Tuckute G, Kauf C, Hosseini EA, Kanwisher N, et al. The neural architecture of language: Integrative modeling converges on predictive processing. Proceedings of the National Academy of Sciences 2021 Nov;118(45):e2105646118. http://www.pnas.org/lookup/doi/10.1073/pnas.2105646118.

[14] Caucheteux C, King JR. Brains and algorithms partially converge in natural language processing. Communications Biology 2022 Dec;5(1):134. https://www.nature.com/articles/s42003-022-03036-1.

[15] Merlin G, Toneva M, Language models and brain alignment: beyond word-level semantics and prediction. arXiv; 2022. http://arxiv.org/abs/2212.00596, arXiv:2212.00596 [cs, q-bio].

[16] Kumar S, Sumers TR, Yamakoshi T, Goldstein A, Hasson U, Norman KA, et al., Reconstructing the cascade of language processing in the brain using the internal computations of a transformer-based language model. bioRxiv; 2022. https://www.biorxiv.org/content/10.1101/2022.06.08.495348v2, pages: 2022.06.08.495348 Section: New Results.

[17] Goldstein A, Zada Z, Buchnik E, Schain M, Price A, Aubrey B, et al. Shared computational principles for language processing in humans and deep language models. Nature Neuroscience 2022 Mar;25(3):369–380. https://www.nature.com/articles/s41593-022-01026-4.

[18] Hosseini EA, Schrimpf M, Zhang Y, Bowman S, Zaslavsky N, Fedorenko E. Artificial neural network language models predict human brain responses to language even after a developmentally realistic amount of training. Neurobiology of Language 2024;5(1):43–63.

[19] Tuckute G, Kanwisher N, Fedorenko E. Language in brains, minds, and machines. Annual Review of Neuroscience 2024;47.

[20] Cadieu CF, Hong H, Yamins DL, Pinto N, Ardila D, Solomon EA, et al. Deep neural networks rival the representation of primate IT cortex for core visual object recognition. PLoS computational biology 2014;10(12):e1003963.

[21] Cichy RM, Khosla A, Pantazis D, Torralba A, Oliva A. Comparison of deep neural networks to spatio-temporal cortical dynamics of human visual object recognition reveals hierarchical correspondence. Scientific Reports 2016 Sep;6(1):27755. http://www.nature.com/articles/srep27755.

[22] Dapello J, Marques T, Schrimpf M, Geiger F, Cox D, DiCarlo JJ. Simulating a Primary Visual Cortex at the Front of CNNs Improves Robustness to Image Perturbations. In: Larochelle H, Ranzato M, Hadsell R, Balcan MF, Lin H, editors. Advances in Neural Information Processing Systems, vol. 33 Curran Associates, Inc.; 2020. p. 13073–13087. https://proceedings.neurips.cc/paper_files/paper/2020/file/98b17f068d5d9b7668e19fb8ae470841-Paper.pdf.

[23] Marques T, Schrimpf M, DiCarlo JJ. Multi-scale hierarchical neural network models that bridge from single neurons in the primate primary visual cortex to object recognition behavior. bioRxiv 2021;p. 2021–03.

[24] Federer C, Xu H, Fyshe A, Zylberberg J. Improved object recognition using neural networks trained to mimic the brain's statistical properties. Neural Networks 2020;131:103–114.

[25] Dapello J, Kar K, Schrimpf M, Geary R, Ferguson M, Cox DD, et al., Aligning Model and Macaque Inferior Temporal Cortex Representations Improves Model-to-Human Behavioral Alignment and Adversarial Robustness. bioRxiv; 2022. https://www.biorxiv.org/content/10.1101/2022.07.01.498495v1, pages: 2022.07.01.498495 Section: New Results.

[26] Toneva M, Wehbe L. Interpreting and improving natural-language processing (in machines) with natural language-processing (in the brain). Advances in neural information processing systems 2019;p. 11.

[27] Schwartz D, Toneva M, Wehbe L, Inducing brain-relevant bias in natural language processing models. arXiv; 2019. http://arxiv.org/abs/1911.03268, arXiv:1911.03268 [cs, q-bio].

[28] Zador A, Escola S, Richards B, Ölveczky B, Bengio Y, Boahen K, et al. Catalyzing next-generation Artificial Intelligence through NeuroAI. Nature Communications 2023 Mar;14(1):1597. https://www.nature.com/articles/s41467-023-37180-x, number: 1 Publisher: Nature Publishing Group.

[29] Schrimpf M, Kubilius J, Lee MJ, Ratan Murty NA, Ajemian R, DiCarlo JJ. Integrative benchmarking to advance neurally mechanistic models of human intelligence. Neuron 2020 Nov;108(3):413–423. https://linkinghub.elsevier.com/retrieve/pii/S089662732030605X.

[30] Chang L, Tsao DY. The code for facial identity in the primate brain. Cell 2017;169(6):1013–1028.

[31] Rudin C, Radin J. Why Are We Using Black Box Models in AI When We Don't Need To? A Lesson From an Explainable AI Competition. Harvard Data Science Review 2019 nov 22;1(2). Https://hdsr.mitpress.mit.edu/pub/f9kuryi8.

[32] Cowley BR, Williamson RC, Acar K, Smith MA, Yu BM. Adaptive stimulus selection for optimizing neural population responses. Advances in neural information processing systems 2017;p. 11.

[33] Bashivan P, Kar K, DiCarlo JJ. Neural population control via deep image synthesis. Science 2019 May;364(6439):eaav9436. https://www.science.org/doi/10.1126/science.aav9436.

[34] Ponce CR, Xiao W, Schade PF, Hartmann TS, Kreiman G, Livingstone MS. Evolving Images for Visual Neurons Using a Deep Generative Network Reveals Coding Principles and Neuronal Preferences. Cell 2019 May;177(4):999–1009.e10. https://linkinghub.elsevier.com/retrieve/pii/S0092867419303915.

[35] Golan T, Siegelman M, Kriegeskorte N, Baldassano C. Testing the limits of natural language models for predicting human language judgements. Nature Machine Intelligence 2023;p. 1–13.

[36] Tuckute G, Sathe A, Srikant S, Taliaferro M, Wang M, Schrimpf M, et al. Driving and suppressing the human language network using large language models. Nature Human Behaviour 2024;8:544–561. https://doi.org/10.1038/s41562-023-01783-7.

[37] Gu Z, Jamison K, Sabuncu MR, Kuceyeski A, Modulating human brain responses via optimal natural image selection and synthetic image generation. arXiv; 2023. http://arxiv.org/abs/2304.09225, arXiv:2304.09225 [q-bio].

[38] Yamins DLK, DiCarlo JJ. Using goal-driven deep learning models to understand sensory cortex. Nature Neuroscience 2016 Mar;19(3):356–365. http://www.nature.com/articles/nn.4244.

[39] Hubel DH, Wiesel TN. Receptive fields of single neurones in the cat's striate cortex. The Journal of Physiology 1959 Oct;148(3):574–591. https://www.ncbi.nlm.nih.gov/pmc/articles/PMC1363130/.

[40] Hubel DH, Wiesel TN. Receptive fields, binocular interaction and functional architecture in the cat's visual cortex. The Journal of Physiology 1962 Jan;160(1):106–154. https://onlinelibrary.wiley.com/doi/10.1113/jphysiol.1962.sp006837.

[41] Felleman DJ, Van Essen DC. Distributed Hierarchical Processing in the Primate Cerebral Cortex. Cerebral Cortex 1991 Jan;1(1):1–47. https://academic.oup.com/cercor/article-lookup/doi/10.1093/cercor/1.1.1.

[42] Logothetis NK, Pauls J, Poggio T. Shape representation in the inferior temporal cortex of monkeys. Current biology 1995;5(5):552–563.

[43] Hung CP, Kreiman G, Poggio T, DiCarlo JJ. Fast readout of object identity from macaque inferior temporal cortex. Science 2005;310(5749):863–866.

[44] Majaj NJ, Hong H, Solomon EA, DiCarlo JJ. Simple learned weighted sums of inferior temporal neuronal firing rates accurately predict human core object recognition performance. Journal of Neuroscience 2015;35(39):13402–13418.

[45] Gross CG, Rocha-Miranda Cd, Bender D. Visual properties of neurons in inferotemporal cortex of the Macaque. Journal of neurophysiology 1972;35(1):96–111.

[46] Tanaka K. Inferotemporal cortex and object vision. Annual review of neuroscience 1996;19(1):109–139.

[47] Tsao DY, Freiwald WA, Tootell RB, Livingstone MS. A cortical region consisting entirely of face-selective cells. Science 2006;311(5761):670–674.

[48] Rajalingham R, DiCarlo JJ. Reversible inactivation of different millimeter-scale regions of primate IT results in different patterns of core object recognition deficits. Neuron 2019;102(2):493–505.

[49] Kar K, DiCarlo JJ. Fast Recurrent Processing via Ventrolateral Prefrontal Cortex Is Needed by the Primate Ventral Stream for Robust Core Visual Object Recognition. Neuron 2021 Jan;109(1):164–176.e5.

[50] Kietzmann TC, Spoerer CJ, Sörensen LK, Cichy RM, Hauk O, Kriegeskorte N. Recurrence is required to capture the representational dynamics of the human visual system. Proceedings of the National Academy of Sciences 2019;116(43):21854–21863.

[51] Sanghavi S, Kar K. Distinct roles of putative excitatory and inhibitory neurons in the macaque inferior temporal cortex in core object recognition behavior. bioRxiv 2023;p. 2023–08.

[52] Kubilius J, Schrimpf M, Kar K, Rajalingham R, Hong H, Majaj N, et al. Brain-Like Object Recognition with High-Performing Shallow Recurrent ANNs. In: Advances in Neural Information Processing Systems 32 (NeurIPS 2019); 2019. https://papers.nips.cc/paper_files/paper/2019/hash/7813d1590d28a7dd372ad54b5d29d033-Abstract.html.

[53] Nayebi A, Bear D, Kubilius J, Kar K, Ganguli S, Sussillo D, et al. Task-Driven Convolutional Recurrent Models of the Visual System. In: Advances in Neural Information Processing Systems 31 (NeurIPS 2018); 2018. https://proceedings.neurips.cc/paper_files/paper/2018/file/6be93f7a96fed60c477d30ae1de032fd-Paper.pdf.

[54] Cornford J, Kalajdzievski D, Leite M, Lamarquette A, Kullmann DM, Richards BA. Learning to live with Dale's principle: ANNs with separate excitatory and inhibitory units. In: International Conference on Learning Representations; 2020. .

[55] Finzi D, Margalit E, Kay K, Yamins DL, Grill-Spector K. A single computational objective drives specialization of streams in visual cortex. bioRxiv 2023;p. 2023–12.

[56] Margalit E, Lee H, Finzi D, DiCarlo JJ, Grill-Spector K, Yamins DLK. A Unifying Principle for the Functional Organization of Visual Cortex. bioRxiv 2023;p. 2023.05.18.541361.

[57] Blauch NM, Behrmann M, Plaut DC. A connectivity-constrained computational account of topographic organization in primate high-level visual cortex. Proceedings of the National Academy of Sciences 2022;119(3):e2112566119.

[58] Dobs K, Martinez J, Kell AJ, Kanwisher N. Brain-like functional specialization emerges spontaneously in deep neural networks. Science advances 2022;8(11):eabl8913.

[59] Seeliger K, Sommers RP, Güçlü U, Bosch SE, Gerven MAJv, A large single-participant fMRI dataset for probing brain responses to naturalistic stimuli in space and time; 2021. https://repository.ubn.ru.nl/handle/2066/228956, accepted: 2021-01-22T23:11:55Z Publisher: Radboud Data Repository.

[60] Khosla M, Wehbe L. High-level visual areas act like domain-general filters with strong selectivity and functional specialization. bioRxiv 2022;p. 2022–03.

[61] Broca P. Remarks on the Seat of the Faculty of Articulated Language, Following an Observation of Aphemia (Loss of Speech). Bulletin de la Société Anatomique 1861;6:330–357. Translation by Christopher D. Green.

[62] Wernicke C. Der Aphasische Symptomencomplex. Cohn and Weigert, Breslau 1874;.

[63] GESCHWIND N. DISCONNEXION SYNDROMES IN ANIMALS AND MAN1. Brain 1965 Jun;88(2):237. https://doi.org/10.1093/brain/88.2.237.

[64] Lichteim L. On Aphasia1. Brain 1885 Jan;7(4):433–484. https://doi.org/10.1093/brain/7.4.433.

[65] Wernicke C. Der aphasische Symptomenkomplex. In: Wernicke C, editor. Der aphasische Symptomencomplex: Eine psychologische Studie auf anatomischer Basis Berlin, Heidelberg: Springer; 1974.p. 1–70. https://doi.org/10.1007/978-3-642-65950-8_1.

[66] Tremblay P, Dick AS. Broca and Wernicke are dead, or moving past the classic model of language neurobiology. Brain and Language 2016 Nov;162:60–71. https://www.sciencedirect.com/science/article/pii/S0093934X16300475.

[67] Fedorenko E, Piantadosi ST, Gibson EA. Language is primarily a tool for communication rather than thought. Nature 2024;630(8017):575–586.

[68] Friederici AD. Towards a neural basis of auditory sentence processing. Trends in Cognitive Sciences 2002;6:78–84. https://api.semanticscholar.org/CorpusID:16634731.

[69] Hickok G, Poeppel D. The cortical organization of speech processing. Nature Reviews Neuroscience 2007 May;8(5):393–402. https://www.nature.com/articles/nrn2113, number: 5 Publisher: Nature Publishing Group.

[70] Friederici AD. The Brain Basis of Language Processing: From Structure to Function. Physiological Reviews 2011 Oct;91(4):1357–1392. https://journals.physiology.org/doi/full/10.1152/physrev.00006.2011, publisher: American Physiological Society.

[71] Fedorenko E, Blank IA, Siegelman M, Mineroff Z. Lack of selectivity for syntax relative to word meanings throughout the language network. Cognition 2020 Oct;203:104348.

[72] Hale J. A Probabilistic Earley Parser as a Psycholinguistic Model. In: Second Meeting of the North American Chapter of the Association for Computational Linguistics; 2001. https://aclanthology.org/N01-1021.

[73] Smith NJ, Levy R. The effect of word predictability on reading time is logarithmic. Cognition 2013 Sep;128(3):302–319. https://www.sciencedirect.com/science/article/pii/S0010027713000413.

[74] Devlin J, Chang MW, Lee K, Toutanova K. BERT: Pre-training of Deep Bidirectional Transformers for Language Understanding. In: Proceedings of NAACL-HLT 2019; 2019. .

[75] Radford A, Narasimhan K, Salimans T, Sutskever I. Improving Language Understanding by Generative Pre-Training; 2018. https://www.semanticscholar.org/paper/Improving-Language-Understanding-by-Generative-Radford-Narasimhan/cd18800a0fe0b668a1cc19f2ec95b5003d0a5035.

[76] Brown TB, Mann B, Ryder N, Subbiah M, Kaplan J, Dhariwal P, et al. Language Models are Few-Shot Learners. arXiv:200514165 [cs] 2020 Jul;http://arxiv.org/abs/2005.14165, arXiv: 2005.14165.

[77] Linzen T, Baroni M. Syntactic Structure from Deep Learning. Annual Review of Linguistics 2021 Jan;7(1):195–212. http://arxiv.org/abs/2004.10827, arXiv:2004.10827 [cs].

[78] Pavlick E. Semantic structure in deep learning. Annual Review of Linguistics 2022;8:447–471.

[79] Chang TA, Bergen BK. Language model behavior: A comprehensive survey. Computational Linguistics 2024;p. 1–58.

[80] Fedorenko E, Thompson-Schill SL. Reworking the language network. Trends in Cognitive Sciences 2014 Mar;18(3):120–126.

[81] Braga RM, DiNicola LM, Becker HC, Buckner RL. Situating the left-lateralized language network in the broader organization of multiple specialized large-scale distributed networks. Journal of Neurophysiology 2020 Nov;124(5):1415–1448.

[82] Lipkin B, Tuckute G, Affourtit J, Small H, Mineroff Z, Kean H, et al. Probabilistic atlas for the language network based on precision fMRI data from >800 individuals. Scientific Data 2022 Aug;9(1):529. https://www.nature.com/articles/s41597-022-01645-3, number: 1 Publisher: Nature Publishing Group.

[83] Fedorenko E, Ivanova AA, Regev TI. The language network as a natural kind within the broader landscape of the human brain. Nature Reviews Neuroscience 2024;p. 1–24.

[84] Kauf C, Tuckute G, Levy R, Andreas J, Fedorenko E. Lexical semantic content, not syntactic structure, is the main contributor to ANN-brain similarity of fMRI responses in the language network. bioRxiv 2023;p. 2023–05.

[85] Deng J, Dong W, Socher R, Li LJ, Li K, Fei-Fei L. ImageNet: A large-scale hierarchical image database. In: 2009 IEEE Conference on Computer Vision and Pattern Recognition; 2009. p. 248–255. ISSN: 1063-6919.

20 TUCKUTE ET AL.[86] Hebart MN, Dickter AH, Kidder A, Kwok WY, Corriveau A, Van Wicklin C, et al. THINGS: A database of 1,854 object concepts and more than 26,000 naturalistic object images. PloS one 2019;14(10):e0223792.

[87] Giverin C, Ramezanpour H, Kar K. Low-cost, portable, easy-to-use kiosks to facilitate home-cage testing of non-human primates during vision-based behavioral tasks. Open Science Framework; 2023.

[88] Rajalingham R, Issa EB, Bashivan P, Kar K, Schmidt K, DiCarlo JJ. Large-scale, high-resolution comparison of the core visual object recognition behavior of humans, monkeys, and state-of-the-art deep artificial neural networks. The Journal of Neuroscience 2018 Aug;38(33):7255–7269. https://www.jneurosci.org/lookup/doi/10.1523/JNEUROSCI.0388-18.2018.

[89] Kell AJ, Bokor SL, Jeon YN, Toosi T, Issa EB. Marmoset core visual object recognition behavior is comparable to that of macaques and humans. Iscience 2023;26(1).

[90] Majaj NJ, Hong H, Solomon EA, DiCarlo JJ. Simple Learned Weighted Sums of Inferior Temporal Neuronal Firing Rates Accurately Predict Human Core Object Recognition Performance. The Journal of Neuroscience: The Official Journal of the Society for Neuroscience 2015 Sep;35(39):13402–13418.

[91] Siegel M, Buschman TJ, Miller EK. Cortical information flow during flexible sensorimotor decisions. Science 2015;348(6241):1352–1355.

[92] Grewe BF, Langer D, Kasper H, Kampa BM, Helmchen F. High-speed in vivo calcium imaging reveals neuronal network activity with near-millisecond precision. Nature methods 2010;7(5):399–405.

[93] Prince JS, Charest I, Kurzawski JW, Pyles JA, Tarr MJ, Kay KN. Improving the accuracy of single-trial fMRI response estimates using GLMsingle. Elife 2022;11:e77599.

[94] Gordon EM, Laumann TO, Gilmore AW, Newbold DJ, Greene DJ, Berg JJ, et al. Precision Functional Mapping of Individual Human Brains. Neuron 2017 Aug;95(4):791–807.e7. https://www.sciencedirect.com/science/article/pii/S089662731730613X.

[95] Gratton C, Braga RM. Editorial overview: Deep imaging of the individual brain: past, practice, and promise. Current Opinion in Behavioral Sciences 2021 Aug;40:iii–vi. https://www.sciencedirect.com/science/article/pii/S2352154621001303.

[96] Smith DM, Perez DC, Porter A, Dworetsky A, Gratton C. Light Through the Fog: Using Precision fMRI Data to Disentangle the Neural Substrates of Cognitive Control. Current Opinion in Behavioral Sciences 2021 Aug;40:19–26.

[97] Allen EJ, St-Yves G, Wu Y, Breedlove JL, Prince JS, Dowdle LT, et al. A massive 7T fMRI dataset to bridge cognitive neuroscience and artificial intelligence. Nature Neuroscience 2022 Jan;25(1):116–126. https://www.nature.com/articles/s41593-021-00962-x, number: 1 Publisher: Nature Publishing Group.

[98] Azadi R, Bohn S, Lopez E, Lafer-Sousa R, Wang K, Eldridge MA, et al. Image-dependence of the detectability of optogenetic stimulation in macaque inferotemporal cortex. Current biology 2023;33(3):581–588.

[99] Rajalingham R, Sorenson M, Azadi R, Bohn S, DiCarlo JJ, Afraz A. Chronically implantable LED arrays for behavioral optogenetics in primates. Nature Methods 2021;18(9):1112–1116.

[100] Afraz SR, Kiani R, Esteky H. Microstimulation of inferotemporal cortex influences face categorization. Nature 2006;442(7103):692–695.

[101] Pascual-Leone A, Gates JR, Dhuna A. Induction of speech arrest and counting errors with rapid-rate transcranial magnetic stimulation. Neurology 1991;41(5):697–702.

[102] Pobric G, Jefferies E, Ralph MAL. Category-specific versus category-general semantic impairment induced by transcranial magnetic stimulation. Current biology 2010;20(10):964–968.

[103] Schalk G, Kapeller C, Guger C, Ogawa H, Hiroshima S, Lafer-Sousa R, et al. Facephenes and rainbows: Causal evidence for functional and anatomical specificity of face and color processing in the human brain. Proceedings of the National Academy of Sciences 2017;114(46):12285–12290.

[104] Simmons JM, Minamimoto T, Murray EA, Richmond BJ. Selective ablations reveal that orbital and lateral prefrontal cortex play different roles in estimating predicted reward value. Journal of Neuroscience 2010;30(47):15878–15887.

[105] El-Shamayleh Y, Horwitz GD. Primate optogenetics: Progress and prognosis. Proceedings of the National Academy of Sciences 2019;116(52):26195–26203.

[106] Roth BL. DREADDs for Neuroscientists. Neuron 2016 Feb;89(4):683–694. https://linkinghub.elsevier.com/retrieve/pii/S0896627316000659.

[107] Fukushima K. Neocognitron: A self-organizing neural network model for a mechanism of pattern recognition unaffected by shift in position. Biological Cybernetics 1980 Apr;36(4):193–202. https://doi.org/10.1007/BF00344251.

[108] Riesenhuber M, Poggio T. Hierarchical models of object recognition in cortex. Nature Neuroscience 1999 Nov;2(11):1019–1025.

[109] Schrimpf M, Kubilius J, Hong H, Majaj NJ, Rajalingham R, Issa EB, et al. Brain-score: which artificial neural network for object recognition is most brain-like? Neuroscience; 2018.

[110] Shannon CE. A Mathematical Theory of Communication. Bell System Technical Journal 1948;27(3):379–423. https://onlinelibrary.wiley.com/doi/abs/10.1002/j.1538-7305.1948.tb01338.x, _eprint: https://onlinelibrary.wiley.com/doi/pdf/10.1002/j.1538-7305.1948.tb01338.x.

[111] Jurafsky D, Martin JH. Speech and Language Processing. 2 ed. Prentice Hall; 2008.

[112] Landauer TK, Dumais ST. A solution to Plato's problem: The latent semantic analysis theory of acquisition, induction, and representation of knowledge. Psychological review 1997;104(2):211.

[113] Mikolov T, Chen K, Corrado G, Dean J. Efficient estimation of word representations in vector space. arXiv preprint arXiv:13013781 2013;.

[114] Pennington J, Socher R, Manning C. Glove: Global Vectors for Word Representation. In: Proceedings of the 2014 Conference on Empirical Methods in Natural Language Processing (EMNLP) Doha, Qatar: Association for Computational Linguistics; 2014. p. 1532–1543. http://aclweb.org/anthology/D14-1162.

[115] Vaswani A, Shazeer N, Parmar N, Uszkoreit J, Jones L, Gomez AN, et al. Attention Is All You Need. In: Advances in Neural Information Processing Systems 30 (NIPS 2017); 2017. .

[116] OpenAI, GPT-4 Technical Report. arXiv; 2023. http://arxiv.org/abs/2303.08774, arXiv:2303.08774 [cs].

[117] Wang A, Singh A, Michael J, Hill F, Levy O, Bowman SR. GLUE: A Multi-Task Benchmark and Analysis Platform for Natural Language Understanding. In: International Conference on Learning Representations; 2019. https://openreview.net/forum?id=rJ4km2R5t7.

[118] Wang A, Pruksachatkun Y, Nangia N, Singh A, Michael J, Hill F, et al. Superglue: A stickier benchmark for general-purpose language understanding systems. Advances in neural information processing systems 2019;32.

[119] Sakaguchi K, Bras RL, Bhagavatula C, Choi Y. Winogrande: An adversarial winograd schema challenge at scale. Communications of the ACM 2021;64(9):99–106.

[120] Kojima T, Gu SS, Reid M, Matsuo Y, Iwasawa Y. Large language models are zero-shot reasoners. Advances in neural information processing systems 2022;35:22199–22213.

[121] Kar K, Kornblith S, Fedorenko E. Interpretability of artificial neural network models in artificial intelligence versus neuroscience. Nature Machine Intelligence 2022;4(12):1065–1067.

[122] Huth AG, de Heer WA, Griffiths TL, Theunissen FE, Gallant JL. Natural speech reveals the semantic maps that tile human cerebral cortex. Nature 2016 Apr;532(7600):453–458. https://www.nature.com/articles/nature17637, number: 7600 Publisher: Nature Publishing Group.

[123] Wang AY, Kay K, Naselaris T, Tarr MJ, Wehbe L. Better models of human high-level visual cortex emerge from natural language supervision with a large and diverse dataset. Nature Machine Intelligence 2023;p. 1–12.

[124] Dharmaretnam D, Foster C, Fyshe A. Words as a window: Using word embeddings to explore the learned representations of Convolutional Neural Networks. Neural Networks 2021;137:63–74.

[125] Hernandez E, Schwettmann S, Bau D, Bagashvili T, Torralba A, Andreas J. Natural language descriptions of deep visual features. In: International Conference on Learning Representations; 2021. .

[126] Güçlü U, van Gerven MA. Deep neural networks reveal a gradient in the complexity of neural representations across the ventral stream. Journal of Neuroscience 2015;35(27):10005–10014.

[127] Kar K, DiCarlo JJ. The Quest for an Integrated Set of Neural Mechanisms Underlying Object Recognition in Primates. arXiv preprint arXiv:231205956 2023;.

[128] Hosseini EA, Schrimpf M, Zhang Y, Bowman S, Zaslavsky N, Fedorenko E, Artificial neural network language models align neurally and behaviorally with humans even after a developmentally realistic amount of training. bioRxiv; 2022. https://www.biorxiv.org/content/10.1101/2022.10.04.510681v1, pages: 2022.10.04.510681 Section: New Results.

[129] Antonello R, Huth A. Predictive Coding or Just Feature Discovery? An Alternative Account of Why Language Models Fit Brain Data. Neurobiology of Language 2023 Feb;p. 1–16. https://doi.org/10.1162/nol_a_00087.

[130] Cohen U, Chung S, Lee DD, Sompolinsky H. Separability and geometry of object manifolds in deep neural networks. Nature communications 2020;11(1):746.

[131] Mamou J, Le H, Del Rio MA, Stephenson C, Tang H, Kim Y, et al. Emergence of Separable Manifolds in Deep Language Representations. In: Proceedings of the 37th International Conference on Machine Learning ICML'20, JMLR.org; 2020. .

[132] Pokorny J, Smith VC. Fifty years exploring the visual system. Annual Review of Vision Science 2020;6:1–23.

[133] Tollefson J. US science agencies on track to hit 25-year funding low. Nature 2023;622(7983):438–439.

[134] Naselaris T, Bassett DS, Fletcher AK, Kording K, Kriegeskorte N, Nienborg H, et al. Cognitive computational neuroscience: a new conference for an emerging discipline. Trends in cognitive sciences 2018;22(5):365–367.

[135] Perrett DI, Oram MW. Neurophysiology of shape processing. Image and Vision Computing 1993;11(6):317–333.

[136] Tang H, Schrimpf M, Lotter W, Moerman C, Paredes A, Ortega Caro J, et al. Recurrent computations for visual pattern completion. Proceedings of the National Academy of Sciences of the United States of America 2018 Aug;115(35):8835–8840.

[137] DiCarlo JJ, Zoccolan D, Rust NC. How does the brain solve visual object recognition? Neuron 2012;73(3):415–434.

[138] Gattass R, Gross CG, Sandell JH. Visual topography of V2 in the macaque. Journal of Comparative Neurology 1981;201(4):519–539.

[139] Gattass R, Sousa A, Gross C. Visuotopic organization and extent of V3 and V4 of the macaque. Journal of neuroscience 1988;8(6):1831–1845.

[140] Op De Beeck H, Vogels R. Spatial sensitivity of macaque inferior temporal neurons. Journal of Comparative Neurology 2000;426(4):505–518.

[141] Kanwisher N, McDermott J, Chun MM. The Fusiform Face Area: A Module in Human Extrastriate Cortex Specialized for Face Perception. Journal of Neuroscience 1997 Jun;17(11):4302–4311. https://www.jneurosci.org/content/17/11/4302, publisher: Society for Neuroscience Section: Articles.

[142] Weiner KS, Sayres R, Vinberg J, Grill-Spector K. fMRI-adaptation and category selectivity in human ventral temporal cortex: regional differences across time scales. Journal of neurophysiology 2010;103(6):3349–3365.

[143] Vinken K, Boix X, Kreiman G. Incorporating intrinsic suppression in deep neural networks captures dynamics of adaptation in neurophysiology and perception. Science Advances 2020 Oct;6(42):eabd4205. https://www.science.org/doi/10.1126/sciadv.abd4205.

[144] Kriegeskorte N. Deep Neural Networks: A New Framework for Modeling Biological Vision and Brain Information Processing. Annual Review of Vision Science 2015 Nov;1:417–446.

[145] Lu Z, Doerig A, Bosch V, Krahmer B, Kaiser D, Cichy RM, et al. End-to-end Topographic Networks as Models of Cortical Map Formation and Human Visual Behavior: Moving Beyond Convolutions. arXiv preprint arXiv:230809431 2023 August;.

[146] Feinman R, Lake BM. Learning inductive biases with simple neural networks. arXiv preprint arXiv:180202745 2018;.

[147] Hasson U, Nastase SA, Goldstein A. Direct fit to nature: an evolutionary perspective on biological and artificial neural networks. Neuron 2020 Feb;105(3):416–434. https://www.sciencedirect.com/science/article/pii/S089662731931044X.

[148] Muttenthaler L, Dippel J, Linhardt L, Vandermeulen RA, Kornblith S. Human alignment of neural network representations. arXiv preprint arXiv:221101201 2022;.

[149] Muttenthaler L, Linhardt L, Dippel J, Vandermeulen RA, Hermann K, Lampinen A, et al. Improving neural network representations using human similarity judgments. Advances in Neural Information Processing Systems 2024;36.

[150] Burg MF, Cadena SA, Denfield GH, Walker EY, Tolias AS, Bethge M, et al. Learning divisive normalization in primary visual cortex. PLOS Computational Biology 2021;17(6):e1009028.

[151] Pirlot C, Gerum RC, Efird C, Zylberberg J, Fyshe A. Improving the Accuracy and Robustness of CNNs Using a Deep CCA Neural Data Regularizer. arXiv preprint arXiv:220902582 2022;.

[152] Kriegeskorte N, Mur M, Bandettini PA. Representational similarity analysis-connecting the branches of systems neuroscience. Frontiers in systems neuroscience 2008;p. 4.

[153] Mishra A, Kanojia D, Nagar S, Dey K, Bhattacharyya P. Leveraging cognitive features for sentiment analysis. arXiv preprint arXiv:170105581 2017;.

[154] Hollenstein N, Barrett M, Troendle M, Bigiolli F, Langer N, Zhang C. Advancing NLP with cognitive language processing signals. arXiv preprint arXiv:190402682 2019;.

[155] Muttenthaler L, Hollenstein N, Barrett M, Human brain activity for machine attention. arXiv; 2020. http://arxiv.org/abs/2006.05113, arXiv:2006.05113 [cs, q-bio].

[156] Ortega Caro J, Oliveira Fonseca AH, Averill C, Rizvi SA, Rosati M, Cross JL, et al. BrainLM: A foundation model for brain activity recordings. bioRxiv 2023;p. 2023–09.

[157] Lurz KK, Bashiri M, Willeke K, Jagadish A, Wang E, Walker EY, et al. Generalization in data-driven models of primary visual cortex. In: International Conference on Learning Representations; 2020. .

[158] Qian WW, Wei JN, Sanchez-Lengeling B, Lee BK, Luo Y, Vlot M, et al. Metabolic activity organizes olfactory representations. Elife 2023;12:e82502.

[159] Pasupathy A, Connor CE. Shape representation in area V4: position-specific tuning for boundary conformation. Journal of Neurophysiology 2001 Nov;86(5):2505–2519.

[160] Movshon JA, Thompson ID, Tolhurst DJ. Spatial summation in the receptive fields of simple cells in the cat's striate cortex. The Journal of physiology 1978;283(1):53–77.

[161] Ringach DL, Hawken MJ, Shapley R. Dynamics of orientation tuning in macaque primary visual cortex. Nature 1997;387(6630):281–284.

[162] Chichilnisky E. A simple white noise analysis of neuronal light responses. Network: computation in neural systems 2001;12(2):199.

[163] Vinje WE, Gallant JL. Sparse coding and decorrelation in primary visual cortex during natural vision. Science 2000;287(5456):1273–1276.

[164] van Hateren Jv, Rüttiger L, Sun H, Lee B. Processing of natural temporal stimuli by macaque retinal ganglion cells. Journal of Neuroscience 2002;22(22):9945–9960.

[165] Nishimoto S, Vu AT, Naselaris T, Benjamini Y, Yu B, Gallant JL. Reconstructing visual experiences from brain activity evoked by natural movies. Current biology 2011;21(19):1641–1646.

[166] Najemnik J, Geisler WS. Optimal eye movement strategies in visual search. Nature 2005;434(7031):387–391.

[167] Friederici AD, Rüschemeyer SA, Hahne A, Fiebach CJ. The role of left inferior frontal and superior temporal cortex in sentence comprehension: localizing syntactic and semantic processes. Cerebral cortex 2003;13(2):170–177.

[168] Hagoort P, Hald L, Bastiaansen M, Petersson KM. Integration of Word Meaning and World Knowledge in Language Comprehension. Science 2004 Apr;304(5669):438–441. https://www.science.org/doi/abs/10.1126/science.1095455, publisher: American Association for the Advancement of Science.

[169] Pylkkänen L, Llinás R, Murphy GL. The representation of polysemy: MEG evidence. Journal of cognitive neuroscience 2006;18(1):97–109.

[170] Henderson JM, Choi W, Lowder MW, Ferreira F. Language structure in the brain: A fixation-related fMRI study of syntactic surprisal in reading. NeuroImage 2016 May;132:293–300.

[171] Garofolo JS. Timit acoustic phonetic continuous speech corpus. Linguistic Data Consortium, 1993 1993;.

[172] Futrell R, Gibson E, Tily HJ, Blank I, Vishnevetsky A, Piantadosi ST, et al. The Natural Stories corpus: A reading-time corpus of English texts containing rare syntactic constructions. Language Resources and Evaluation 2021;55(1):63–77.

[173] Hamilton LS, Huth AG. The revolution will not be controlled: natural stimuli in speech neuroscience. Language, cognition and neuroscience 2020;35(5):573–582.

[174] Djambazovska S, Zafer A, Ramezanpour H, Kreiman G, Kar K. The Impact of Scene Context on Visual Object Recognition: Comparing Humans, Monkeys, and Computational Models. bioRxiv 2024;https://www.biorxiv.org/content/early/2024/06/01/2024.05.27.596127.

[175] Bresnan J, Kanerva JM. Locative inversion in Chicheŵa: A case study of factorization in grammar. Linguistic inquiry 1989;p. 1–50.

[176] Hu J, Gauthier J, Qian P, Wilcox E, Levy RP, A Systematic Assessment of Syntactic Generalization in Neural Language Models. arXiv; 2020. http://arxiv.org/abs/2005.03692, arXiv:2005.03692 [cs].

[177] Geisler WS. Visual perception and the statistical properties of natural scenes. Annu Rev Psychol 2008;59:167–192.

[178] Mehrer J, Spoerer CJ, Jones EC, Kriegeskorte N, Kietzmann TC. An ecologically motivated image dataset for deep learning yields better models of human vision. Proceedings of the National Academy of Sciences 2021;118(8):e2011417118.

[179] Barlow HB, et al. Possible principles underlying the transformation of sensory messages. Sensory communication 1961;1(01):217–233.

[180] Maunsell J, Nealey TA, DePriest DD. Magnocellular and parvocellular contributions to responses in the middle temporal visual area (MT) of the macaque monkey. Journal of Neuroscience 1990;10(10):3323–3334.

[181] Maunsell JH, Ghose GM, Assad JA, Mcadams CJ, Boudreau CE, Noerager BD. Visual response latencies of magnocellular and parvocellular LGN neurons in macaque monkeys. Visual neuroscience 1999;16(1):1–14.

[182] Blank I, Balewski Z, Mahowald K, Fedorenko E. Syntactic processing is distributed across the language system. NeuroImage 2016 Feb;127:307–323.

[183] Siegelman M, Blank IA, Mineroff Z, Fedorenko E. An Attempt to Conceptually Replicate the Dissociation between Syntax and Semantics during Sentence Comprehension. Neuroscience 2019 Aug;413:219–229.

[184] Caucheteux C, Gramfort A, King JR. Disentangling syntax and semantics in the brain with deep networks. In: Meila M, Zhang T, editors. Proceedings of the 38th International Conference on Machine Learning, vol. 139 of Proceedings of Machine Learning Research PMLR; 2021. p. 1336–1348. https://proceedings.mlr.press/v139/caucheteux21a.html.

[185] Rust NC, Mante V, Simoncelli EP, Movshon JA. How MT cells analyze the motion of visual patterns. Nature neuroscience 2006;9(11):1421–1431.

[186] Ratan Murty NA, Bashivan P, Abate A, DiCarlo JJ, Kanwisher N. Computational models of category-selective brain regions enable high-throughput tests of selectivity. Nature Communications 2021 Sep;12(1):5540. https://www.nature.com/articles/s41467-021-25409-6, number: 1 Publisher: Nature Publishing Group.

[187] Golan T, Raju PC, Kriegeskorte N. Controversial stimuli: Pitting neural networks against each other as models of human cognition. Proceedings of the National Academy of Sciences 2020;117(47):29330–29337.

[188] Benda J, Gollisch T, Machens CK, Herz AV. From response to stimulus: adaptive sampling in sensory physiology. Current opinion in neurobiology 2007;17(4):430–436.

[189] Lewi J, Butera R, Paninski L. Sequential optimal design of neurophysiology experiments. Neural computation 2009;21(3):619–687.

[190] DiMattina C, Zhang K. Adaptive stimulus optimization for sensory systems neuroscience. Frontiers in Neural Circuits 2013;7. http://journal.frontiersin.org/article/10.3389/fncir.2013.00101/abstract.

[191] Pillow JW, Park M. Adaptive Bayesian methods for closed-loop neurophysiology. Closed loop neuroscience 2016;p. 3–18.

[192] Tuckute G, Hansen ST, Kjaer TW, Hansen LK. Real-time decoding of attentional states using closed-loop EEG neurofeedback. Neural Computation 2021;33(4):967–1004.

[193] Mukherjee K, Kim NY, Alamooti ST, Adolphs R, Kar K. Leveraging Artificial Neural Networks to Enhance Diagnostic Efficiency in Autism Spectrum Disorder: A Study on Facial Emotion Recognition. Annual meeting of Cognitive Computational Neuroscience Conference;.

[194] Kriegeskorte N, Mur M, Bandettini P. Representational similarity analysis - connecting the branches of systems neuroscience. Frontiers in Systems Neuroscience 2008;2. https://www.frontiersin.org/articles/10.3389/neuro.06.004.2008.

[195] Kriegeskorte N, Kievit RA. Representational geometry: integrating cognition, computation, and the brain. Trends in Cognitive Sciences 2013 Aug;17(8):401–412.

[196] Kornblith S, Norouzi M, Lee H, Hinton G, Similarity of Neural Network Representations Revisited. arXiv; 2019. http://arxiv.org/abs/1905.00414, arXiv:1905.00414 [cs, q-bio, stat].

[197] Williams AH, Kunz E, Kornblith S, Linderman SW. Generalized Shape Metrics on Neural Representations. In: Advances in Neural Information Processing Systems 34 (NeurIPS 2021); 2021. ArXiv:2110.14739 [cs, stat].

[198] Kolling C, Speicher T, Nanda V, Toneva M, Gummadi KP. Pointwise Representational Similarity. arXiv preprint arXiv:230519294 2023;.

[199] Khosla M, Williams AH. Soft Matching Distance: A metric on neural representations that captures single-neuron tuning. arXiv preprint arXiv:231109466 2023;.

[200] Sexton NJ, Love BC. Reassessing hierarchical correspondences between brain and deep networks through direct interface. Science Advances 2022 Jul;8(28):eabm2219. https://www.science.org/doi/10.1126/sciadv.abm2219, publisher: American Association for the Advancement of Science.

[201] Beguš G, Zhou A, Zhao TC. Encoding of speech in convolutional layers and the brain stem based on language experience. Scientific Reports 2023 Apr;13(1):6480. https://www.nature.com/articles/s41598-023-33384-9, number: 1 Publisher: Nature Publishing Group.

[202] Cichy RM, Roig G, Andonian A, Dwivedi K, Lahner B, Lascelles A, et al. The algonauts project: A platform for communication between the sciences of biological and artificial intelligence. arXiv preprint arXiv:190505675 2019;.

[203] Gifford AT, Lahner B, Saba-Sadiya S, Vilas MG, Lascelles A, Oliva A, et al. The algonauts project 2023 challenge: How the human brain makes sense of natural scenes. arXiv preprint arXiv:230103198 2023;.

[204] Willeke KF, Fahey PG, Bashiri M, Pede L, Burg MF, Blessing C, et al. The Sensorium competition on predicting large-scale mouse primary visual cortex activity. arXiv preprint arXiv:220608666 2022;.